# High-resolution and broadband all-fiber spectrometers

Brandon Redding[1], Mansoor Alam[2], Martin Seifert[2], and Hui Cao[1]*

[1]Department of Applied Physics, Yale University, New Haven CT 06511
[2]Nufern, East Granby CT 06026
*hui.cao@yale.edu

**Abstract**

The development of optical fibers has revolutionized telecommunications by enabling long-distance broad-band transmission with minimal loss. In turn, the ubiquity of high-quality low-cost fibers enabled a number of additional applications, including fiber sensors, fiber lasers, and imaging fiber bundles. Recently, we showed that a multimode optical fiber can also function as a spectrometer by measuring the wavelength-dependent speckle pattern formed by interference between the guided modes. Here, we reach a record resolution of 1 pm at wavelength 1500 nm using a 100 meter long multimode fiber, outperforming the state-of-the-art grating spectrometers. We also achieved broad-band operation with a 4 cm long fiber, covering 400 nm – 750 nm with 1 nm resolution. The fiber spectrometer, consisting of the fiber which can be coiled to a small volume and a monochrome camera that records the speckle pattern, is compact, lightweight, and low cost while providing ultrahigh resolution, broad bandwidth and low loss.

**Introduction**

Spectrometers are widely used tools in chemical and biological sensing, material analysis, and light source characterization. Traditional spectrometers use a grating to disperse light, and the spectral resolution scales with the optical pathlength from the grating to the detectors, imposing a trade-off between device size and resolution. In recent years the development of "miniature" spectrometers has enabled a host of new applications due to their reduced cost and portability. However, they still rely on grating dispersion, thus the spectral resolution is lower than the large bench-top spectrometers. In order to develop a compact spectrometer without sacrificing resolution we turn to a multimode fiber, where a long pathlength is easily achieved in a small footprint by coiling the fiber. Of course, replacing the grating with a multimode fiber also requires the spectrometer to operate in a different paradigm.

In a grating-based spectrometer, light from different spectral bands is mapped to different spatial positions. However, this one-to-one spectral-to-spatial mapping is not strictly required, and spectrometers have also been built using dispersive elements with more complex spectral-to-spatial mapping. In these implementations, different spectral bands are mapped to different spatial intensity patterns. These intensity patterns function as fingerprints identifying the spectral content of an unknown probe. This approach has enabled spectrometers based on disordered photonic crystals [1] and random scattering media[2,3], which have the advantage of being very compact. Multiple scattering of light in a disordered structure increases the optical pathlength by folding the paths in a confined geometry, enhancing the spectral decorrelation of transmitted speckle. The resulting high spatial-spectral diversity enabled the development of a random spectrometer on a silicon chip, where the footprint is especially limited[3]. However, light transmission through a strongly scattering medium is usually low, limiting the sensitivity to weak signals. A photonic bandgap fiber bundle has also been used as a spectrometer[4]. To increase the intensity throughput, the transmission bands of the constituent fibers are broad and overlapping, which limits the wavelength resolution to ~ 10 nm[4].

In this work, we show that using a single multimode fiber as the dispersive element overcomes many of these limitations. In a multimode fiber, interference among the guided modes creates wavelength-dependent speckle patterns, providing the required spectral-to-spatial mapping[5,6]. The spectral resolution is then determined by the minimal change in wavelength that generates an uncorrelated speckle pattern. In a multimode fiber, the spectral correlation width of speckle scales inversely with the length of the fiber[7–9]. Since optical fibers have been optimized for long-distance transmission with minimal loss, long fibers can be used to provide fine resolution without sacrificing sensitivity. In addition, the fiber can be coiled into a small footprint, enabling a compact, high-resolution, low-loss spectrometer.

**High Spectral Resolution Fiber Spectrometer**

Large core optical fibers can easily support hundreds to thousands of spatial modes. The speckle pattern produced by interference between these modes is determined by their relative amplitude and phase. For a monochromatic input light, the electric field at the end of a fiber of length $L$ can be written as the sum of the contribution from each guided mode:

$$\mathbf{E}(r,\theta,\lambda,L) = \sum_m A_m \mathbf{\psi}_m(r,\theta,\lambda) \exp\left[-i\left(\beta_m(\lambda)L - \omega t + \varphi_m\right)\right] \quad (1)$$

where $A_m$ and $\varphi_m$ are the amplitude and initial phase of the $m^{th}$ mode which has spatial profile $\mathbf{\psi_m}$ and propagation constant $\beta_m$. A shift of the input wavelength $\lambda$ modifies the propagation constant causing the guided modes to accumulate different phase delays, $\beta_m(\lambda)L$, as they travel along the fiber, and thereby changing the speckle pattern. The sensitivity of the speckle pattern to a change in wavelength therefore increases with $L$. The spectral resolution of the fiber spectrometer, given by the minimum wavelength shift that produces an uncorrelated speckle pattern, thus scales inversely with the length of the fiber[5].

We note that Eq. 1 assumes an ideal multimode optical fiber in which the modal distribution (i.e. the coefficients $A_m$) remains fixed for the length of the fiber. Experimentally, small imperfections in the fiber or bending/twisting of the fiber introduces mode coupling which can affect the spectral resolution of the fiber spectrometer. In the limit of strong mode coupling, the spectral resolution scales as $L^{-1/2}$ instead of $L^{-1}$. However, the mode coupling length of standard glass core fibers is much longer than the lengths of the fibers used for the spectrometers[10]. Moreover, we observed that the spectral correlation width scaled linearly with the fiber length up to 100 meters, confirming that mode coupling has a negligible effect on the spectral resolution of the fiber spectrometers.

In order to push the resolution limit of the fiber spectrometer, we selected a 100-meter long, step-index multimode fiber (core diameter = 105 µm, NA = 0.22). The fiber supported ~1000 modes at $\lambda$ = 1500 nm[11]. The 100 m fiber was coiled on a 3" diameter spool and fixed by resin for mechanical stability. A photograph of the coiled fiber, placed on top of an insulating block, is shown in Fig. 1(a). A monochrome camera was used to record the far-field speckle patterns. In general, the speckle from a multimode fiber depends not only on the input wavelength, but also on the polarization and spatial profile of the input light. To ensure that the same wavelength always produced identical speckle pattern, the probe light was coupled through a polarization-maintaining single-mode fiber to the multimode fiber. The single mode fiber was fused to the multimode fiber, as seen in Fig. 1(b), to ensure the same combination of guided modes in the multimode fiber was excited every time.

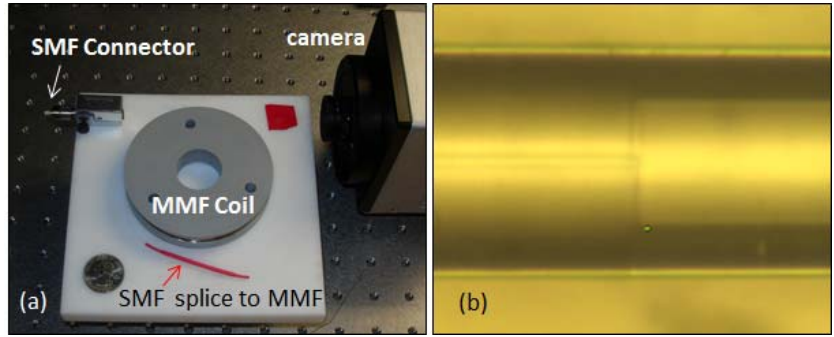

**Figure 1 | 100 meter multimode fiber spectrometer.** (a) A photograph of the fiber spectrometer. Probe light is coupled to the polarization-maintaining single-mode fiber (SMF) which is fused to the multimode fiber (MMF). The light then excites a superposition of guided modes in the multimode fiber, which travel 100 meters around the 3" coil. The output end of the fiber was secured to an insulating block placed in front of a monochrome camera, which records the wavelength-dependent speckle pattern in the far field. (b) Optical microscope image of a single-mode fiber spliced to a multimode fiber, confirming that the single-mode fiber core was fused to the multimode fiber core to minimize the coupling loss. By coupling the probe light through the polarization-maintaining single-mode fiber to the multimode fiber, we ensure that the same combination of guided modes is always excited in the multimode fiber.

A representative speckle pattern, recorded at an input wavelength of $\lambda = 1500$ nm, is shown in Fig. 2(b). Due to the long length of the fiber, a tiny change in wavelength was sufficient to produce uncorrelated speckle patterns. To quantify this sensitivity, we calculated the spectral correlation function of the speckle intensity, $C(\Delta\lambda, x) = \langle I(\lambda,x)I(\lambda+\Delta\lambda,x)\rangle / [\langle I(\lambda,x)\rangle\langle I(\lambda+\Delta\lambda,x)\rangle] - 1$, where $I(\lambda,x)$ is the intensity at a position $x$ for input wavelength $\lambda$, and $\langle...\rangle$ represents the average over $\lambda$. We then averaged $C(\Delta\lambda,x)$ over each position $x$ in the speckle pattern. The correlation function is plotted in Fig. 2(b). The correlation width, defined as the change in wavelength required to reduce the correlation by half, was found to be merely 1.5 pm. This value provides an estimate for the resolution of the fiber spectrometer, although the actual resolution depends also on the noise of the measurements and the reconstruction algorithm.

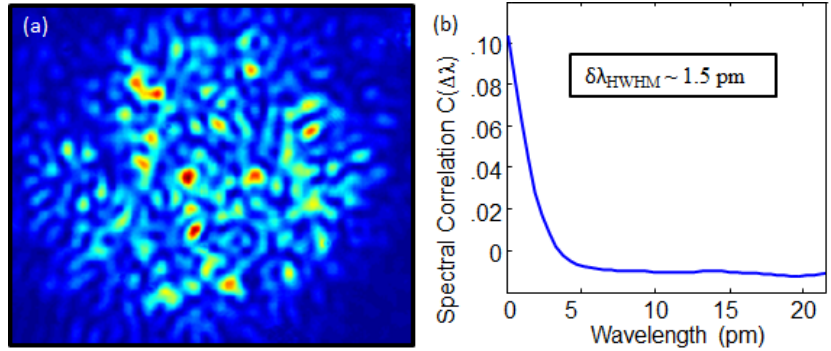

**Figure 2 | Wavelength-dependent speckle pattern from a 100 meter multimode fiber.** (a) A speckle pattern recorded at $\lambda$ = 1500 nm from the end of a 100m-long step-index multimode fiber (core diameter = 105 μm, NA = 0.22). From the number of speckles, we estimated that approximately 400 spatial modes were excited in the multimode fiber. (b) Measured spectral correlation function of speckle, showing a half-width-at-half-maximum (HWHM) of 1.5pm. The small spectral correlation width enables high resolution.

In order to use the 100 meter multimode fiber as a spectrometer, we first calibrated its transmission matrix, which characterizes the spectral-to-spatial mapping properties of the fiber. The transmission matrix, $T$, was defined as $I=T·S$, where $I$ is a vector describing the discretized speckle pattern and $S$ is the discretized input spectra[6]. Each column in $T$ represents the speckle pattern produced by one spectral channel. The calibration procedure consisted of recording the speckle pattern for each spectral channel in $S$, thereby measuring $T$ one column at a time. For the 100 meter fiber spectrometer, we calibrated a transmission matrix consisting of 200 spectral channels from $\lambda$=1500.0 nm to $\lambda$=1500.1 nm in steps of 0.5 pm. The speckle pattern in Fig. 2(a) contains approximately 400 speckles, indicating that ~400 spatial modes were excited in the multimode fiber. We thus selected 400 independent spatial channels for the calibration of $T$. After calibration, we recorded the speckle pattern for various probe wavelengths and attempted to reconstruct the input spectra. The reconstruction was performed using a combination of matrix pseudo-inversion and non-linear optimization[6]. Figure 3(a) shows the reconstructed spectra for a series of narrow spectral lines across the operating bandwidth. The linewidth is ~1 pm and the signal-to-noise ratio (peak-to-background ratio) is over 1000. We then tested the spectral resolution of the 100 meter fiber spectrometer by resolving two closely spaced spectral lines. In order to synthesize such a test probe spectrum, we separately recorded speckle patterns at two closely spaced probe wavelengths and then added these speckle patterns in intensity. We then attempted to reconstruct the spectra from the synthesized speckle pattern. As shown in Fig. 3(b), the fiber spectrometer was able to resolve two lines separated by merely 1 pm. This spectral resolution is beyond the capabilities of even the largest bench-top grating spectrometers, which cannot realistically compete with the long pathlength achieved in an optical fiber.

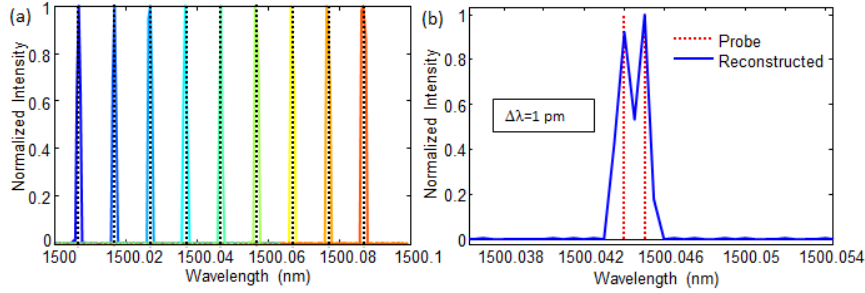

**Figure 3 | Ultra-high resolution of 100 meter fiber spectrometer.** (a) Reconstructed spectra for a series of narrow lines using the 100m-long multimode fiber spectrometer. The probe frequencies are marked by vertical black dotted lines. (b) Reconstructed spectrum (blue line) of two narrow spectral lines separated by 1 pm. The red dotted lines mark the center wavelengths of the probe lines.

Of course, a multimode fiber with such high spectral resolution is also sensitive to environmental perturbations. In fact, multimode fibers have been used for a variety of sensing applications, including temperature[12,13], pressure[14], and vibration sensing[15]. In these systems, a perturbation such as a change in the ambient temperature introduces a slight change in the refractive index of the glass fiber[13]. As a result, light travelling in the fiber experiences slightly different propagation constant. In a sufficiently long fiber, this difference can add up to a significant change of the phase delay and alter the speckle pattern. Since our fiber spectrometer relies on a given wavelength consistently producing the same speckle pattern, changes in temperature would corrupt the reconstruction process.

In order to minimize mechanical perturbations, the fiber was secured to the spool using resin. We also spliced a single mode fiber to the multimode fiber to avoid any change in the input coupling condition [Fig. 1(b)]. However, changes in the ambient temperature in the lab still affected the fiber spectrometer. We estimated that a change of ~0.01° C was sufficient to produce an uncorrelated speckle pattern in a 100 meter fiber [6]. In other words, the fiber environment should be held at a constant temperature, to within ~0.01° C. Although thermal stabilization at this level is possible, it would add complexity and cost to the fiber spectrometer. Next, we present a simple correction technique which can significantly improve the robustness of the fiber spectrometer to temperature fluctuations.

As the ambient temperature varies, the refractive index of the fiber glass changes due to the thermo-optic effect, $n(T+\delta T) = n(T)(1+7\times10^{-6}\,\delta T)$, and the length of the fiber also changes due to thermal expansion, $L(T+\delta T) = L(T)(1+5\times10^{-7}\,\delta T)$[13]. To understand the effect such changes have on the speckle pattern, we refer back to Eq. 1, which expresses the field at the output end of the fiber. The change in the refractive index affects the spatial profile $\psi_m$ of the modes, their amplitude $A_m$ and initial phase $\varphi_m$, as well as the propagation constants $\beta_m$. The resulting change in the accumulated phase, $\beta_m(\lambda,T)L(T)$, is amplified by the length of the fiber. For long fibers and small changes in temperature, the effects of the changes to $\psi_m$, $A_m$ and $\varphi_m$ on the speckle pattern are negligible compared with that to $\beta_m(\lambda,T)L$. Thus we consider only the change in accumulated phase below.

If the temperature changes from $T$ (the temperature during the calibration of the fiber spectrometer) to $T+\delta T$, the phase delay for the $m^{th}$ mode becomes $\beta_m(\lambda,T+\delta T)L(T+\delta T)$. It is equal to the phase delay at $T$ for a slightly different wavelength $\lambda+\delta\lambda$, i.e, $\beta_m(\lambda+\delta\lambda,T)L(T)=\beta_m(\lambda,T+\delta T)L(T+\delta T)$. In principle, the wavelength change $\delta\lambda$ depends on the mode index $m$ in the multimode fiber. However, our numerical simulation reveals that for small temperature variations, the change in phase delay is approximately the same for all modes, thus $\delta\lambda$ is nearly mode-independent (Supplementary Section S1). This means the speckle pattern generated by input light of wavelength $\lambda$ at temperature $T+\delta T$ resembles the speckle pattern produced by the wavelength $\lambda+\delta\lambda$ at temperature $T$. Thus the reconstruction algorithm would recover an input wavelength of $\lambda+\delta\lambda$ from the speckle pattern for $\lambda$ at $T+\delta T$. This allows us to correct the temperature change $\delta T$ by a wavelength shift $\delta\lambda$. Although the value of $\delta\lambda$ depends on $\lambda$, the operating bandwidth (0.1 nm) is much less than the wavelength (1500 nm), and the thermo-optic coefficient and thermal expansion coefficient are almost constant across this bandwidth. Thus the variation of $\delta\lambda$ across the bandwidth is negligible, and we can adjust for a small change in the ambient temperature by applying a single $\delta\lambda$ correction factor. Usually the temperature change is slow, therefore we propose to periodically interrogate the fiber spectrometer with a laser of a fixed wavelength to measure the $\delta\lambda$ that accounts for the current temperature, and then apply this simple correction to subsequent measurements.

To gauge the efficacy of this approach, we experimentally investigated the stability of the fiber spectrometer. Note that the measurements in Fig. 3 were performed within a few minutes after calibrating the multimode fiber. The successful spectral reconstruction indicates that within this time, environmental conditions had not changed significantly. We then continued to record speckle patterns for ~10 hours after calibration. Every ~1.5 minutes, we recorded speckle patterns at 9 wavelengths across the spectrometer operating bandwidth. Figure 4(a) plots the reconstructed spectra from three sets of measurements. In the first set (bottom row), which was recorded ~1 minute after calibration, each probe wavelength was accurately reconstructed. However, in the second and third set of measurements, recorded 10 and 20 minutes after calibration, the reconstructed lines are shifted relative to the probe lines, due to changes in the ambient temperature. Nonetheless, the reconstruction algorithm still identified a narrow line for each measurement. This indicates that the speckle pattern produced by a wavelength $\lambda$ at temperature $T+\delta T$ matched the speckle pattern produced by a nearby wavelength, $\lambda+\delta\lambda$, at the calibration temperature $T$. In addition, we note that for a given set of measurements recorded within a ~30 second time period [i.e. one row in Fig. 4(a)], each reconstructed line was shifted by the same amount. This confirmed our expectation that a single wavelength shift $\delta\lambda$ could be used to correct for the temperature induced changes in all reconstructed spectra.

To correct for this wavelength shift, we assumed that the measurement at $\lambda$ = 1500.045 nm was performed using a known reference laser. We then calculated the $\delta\lambda$ shift between 1500.045 nm and the peak wavelength of the reconstructed spectrum, and applied this same $\delta\lambda$ to the spectra reconstructed from other measurements. As shown in Fig. 4(b), each reconstructed line matched the probe line after this simple correction. To quantitatively assess the efficacy of this correction, we calculated the error between the reconstructed peak wavelength and the actual probe wavelength for all 4000 measurements recorded over the 10 hour testing period. We then repeated the correction procedure by using the first probe line as a reference laser to estimate the current $\delta\lambda$. Figure 4(c) presents a histogram of the errors in the reconstructed wavelengths before and after this correction. Due to thermal fluctuations, the peak position of the reconstructed lines varied by as much as 15 - 20 pm from the actual wavelength before correction. However, after this simple correction, 95% of the measurements were accurate to within ±3 pm. In practice, a reference laser with a well-defined frequency may be used to repeatedly re-calibrate the fiber spectrometer, significantly reducing the sensitivity to environmental perturbations. Note that this software correction was performed without any effort to stabilize the ambient temperature of the multimode fiber. By combining the software correction with thermal and mechanical stabilization, we expect the multimode fiber spectrometer with ultrahigh resolution could be made robust against environmental perturbations.

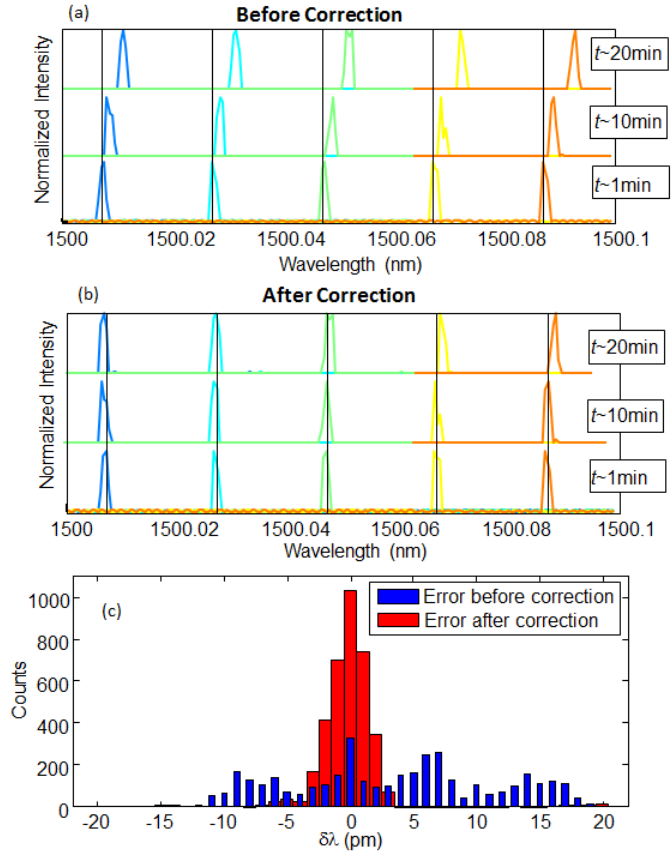

**Figure 4 | Correction of the effects of temperature fluctuation on fiber spectrometer.** (a) Each row represents five spectra reconstructed from the speckle patterns of 5 probe lines measured within ~30 seconds of each other at some time, $t$, after calibration (indicated next to the curves). Vertical black lines mark the spectral position of the probe lines. At $t \sim 1$ min (bottom row), the MMF spectrometer accurately recovers all five probe lines; but at $t \sim 10$ min (middle row) and 20 min (top row), the reconstructed lines shift from the actual wavelengths due to temperature fluctuations. Fortunately, all five lines in the same row (recorded within 30 seconds of each other) drift together, thus a single wavelength shift $\delta\lambda$ can be used to correct for the drift. (b) The wavelength shift $\delta\lambda$ of the probe line at $\lambda = 1500.045$ nm was used to correct the drift of other lines in the same row. The corrected spectra match the probe lines well. (c) Histogram of the error between the input wavelength and reconstructed wavelength, $\delta\lambda$, for 4,000 measurements recorded over 10 hours. Before correction, the error ranges from −15 to +15 pm due to thermal

fluctuations. After correction, 95% of the measurements fall within ±3 pm of the correct wavelength.

**Broadband Fiber Spectrometer**

In addition to ultra-fine spectral resolution, multimode fiber spectrometers can also operate with very broad bandwidth. In this section, we demonstrate broadband operation in the visible spectrum using a 4 cm long multimode fiber (105 μm diameter core, NA = 0.22). In order to calibrate the wavelength dependent speckle patterns in the visible spectrum, we used a broadband supercontinuum light source in combination with a monochromator. A schematic of the experimental setup is provided in the Supplementary Information. The monochromator selected a narrow (~0.2 nm full width at half maximum) band of the supercontinuum emission which was first coupled to a single-mode polarization-maintaining fiber and then to the multimode fiber. The single mode fiber ensured excitation of the same superposition of guided modes in the multimode fiber and allowed us to test different light sources after calibration. To illustrate the speckle patterns formed by different colors, we used a color charge-coupled-device (CCD) camera to record red, green and blue speckles at the end of the multimode fiber, as shown in Fig. 5(a). In the actual calibration and testing of the fiber spectrometer, a monochrome CCD camera was used. From the number of speckles, we estimated that ~700 spatial modes were excited in the multimode fiber.

For the 4 cm fiber, the spectral correlation function of speckle has a HWHM of ~2 nm (see Supplementary Information). We calibrated a transmission matrix consisting of 700 spectral channels from $\lambda$ = 400 nm to $\lambda$ = 750 nm in steps of 0.5 nm, and 4800 spatial channels. Not all the spatial channels are independent, although the shorter wavelengths provide more independent spatial channels than at $\lambda$ = 1500 nm. We oversampled spatially in order to have a better reconstruction of broadband spectra[6]. After calibration, we switched to the lowest resolution grating (150 grooves/mm) in the monochromator and increased the exit slit width to obtain a tunable probe of ~10 nm bandwidth. We recorded a series of speckle patterns across the visible spectrum (400 nm – 750 nm), and separately measured the probe spectrum with a commercial grating spectrometer. The reconstructed spectra, shown in Fig. 5(b), match well the spectra recorded by the grating spectrometer. We also tested the spectral resolution of the 4 cm fiber spectrometer, and found that it was able to resolve two narrow lines separated by 1 nm (see Supplementary Information). Furthermore, unlike the 100 meter fiber spectrometer, the short fiber spectrometer is relatively insensitive to environmental perturbations. For the same change in temperature, the change in accumulated phase delay is reduced by more than 3 orders of magnitude due to the difference in the fiber lengths. Consequently, the 4 cm fiber spectrometer is robust against temperature fluctuation within ~10° C[6].

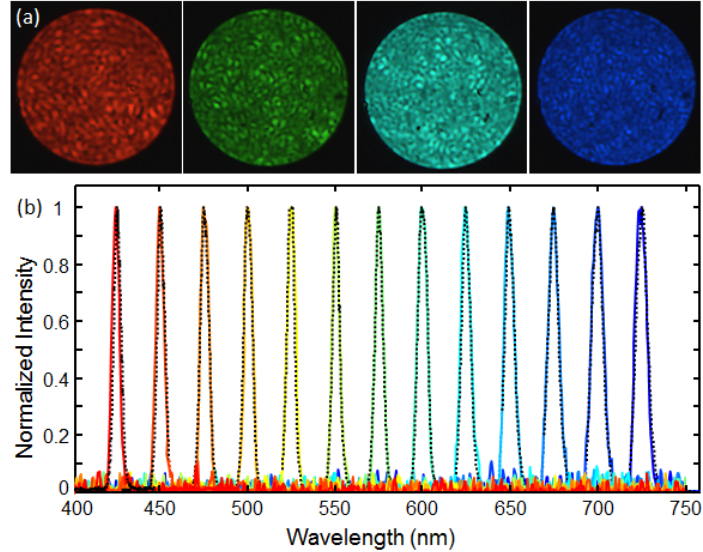

**Figure 5 | Broadband fiber spectrometer.** (a) Speckle patterns recorded using a color CCD camera at the end of a 4 cm multimode fiber. (b) Reconstructed spectra for a series of ~10 nm wide probes across the visible spectrum. The probe spectra measured separately by a grating spectrometer are shown in black dotted lines.

After calibration we used the 4 cm fiber spectrometer to measure the spectra of photoluminescence from Rhodamine 640 dye solution in a cuvette. Figure 6(a) is a schematic of the experimental setup. The output beam from a diode laser operating at $\lambda$=532 nm was focused to the solution to excite the Rhodamine molecules. The emission was collected and focused to a single-mode fiber, which was coupled to the multimode fiber. A long-pass filter was placed in front of the single-mode fiber to block the pump light. The inset of Fig. 6(b) is part of the speckle pattern produced by the photoluminescence at the end of the 4 cm multimode fiber. Despite the broad bandwidth of the emission, speckle was still clearly visible, enabling the fiber spectrometer to reconstruct the spectrum. In Fig. 6(b), the photoluminescence spectrum measured by the fiber spectrometer coincides with the spectrum recorded separately by the grating spectrometer. The accurate measurement of the photoluminescence spectrum also confirmed that the single mode fiber delivered the signal to the multimode fiber in the same spatial mode used in the calibration.

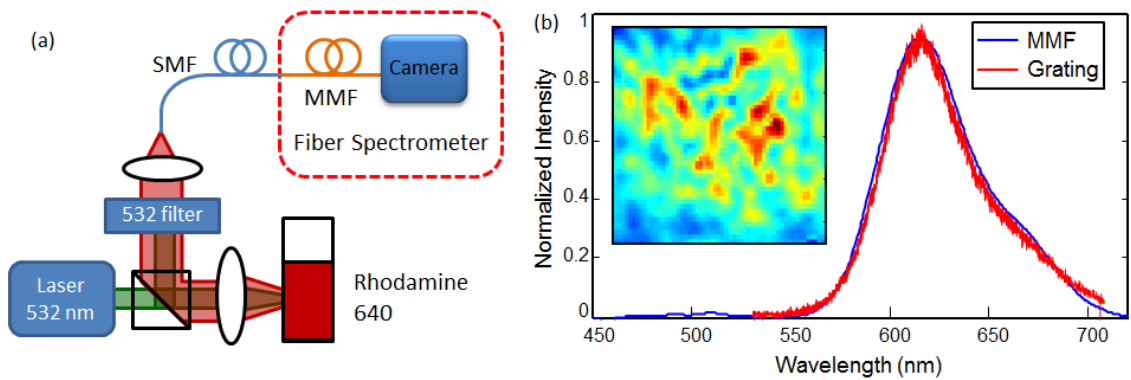

**Figure 6. | Measuring photoluminescence spectra with a fiber spectrometer.** (a) A

schematic of the experimental setup for the photoluminescence measurement. The Rhodamine dye solution in the cuvette was pumped by a diode laser at $\lambda = 532$ nm, and the emission was collected by a single-mode fiber (SMF) which was coupled to the multimode fiber (MMF). (b) Photoluminescence spectrum of the Rhodamine dye measured by the fiber spectrometer (blue solid line) and a grating spectrometer (red dotted line). The inset is a close-up of the speckle produced by the Rhodamine emission at the end of the 4 cm MMF.

**Discussion**

In summary, a multimode fiber spectrometer, consisting of a fiber and a camera, can provide high spectral resolution and large bandwidth while maintaining a small footprint. Using a 100 meter long fiber coiled on a 3" spool, we were able to resolve two lines separated by 1 pm at $\lambda = $ 1500 nm. A simple wavelength correction technique was developed to dramatically improve the stability of the ultrahigh-resolution fiber spectrometer against ambient temperature fluctuation. In the other extreme, a 4 cm long fiber can provide broadband operation, covering the visible spectrum (400 nm – 750 nm) with decent resolution (1 nm). It accurately captured the photoluminescence spectrum of Rhodamine dye.

Multimode fiber spectrometers offer clear advantages over traditional grating spectrometers. The most attractive feature is the ability to achieve high resolution with a compact size. This could enable portable miniature spectrometers with resolution currently only available in large bench-top systems. In addition, optical fiber is extremely low cost, light weight, and has almost negligible loss over the lengths suitable for the spectrometer application. The reconstruction algorithm is fast and provides comparable accuracy to a grating spectrometer[16]. Although high-resolution fiber spectrometers are sensitive to environmental perturbations, a combination of thermal and mechanical stabilization with software correction could enable robust performance.

The main limitation of the fiber spectrometer is that the probe signal must be confined to a fixed spatial mode and polarization state to ensure that a given wavelength always generates the same speckle pattern. In our implementation, this was done by first coupling the probe signal to a single-mode polarization-maintaining fiber. This is analogous to the use of an entrance slit in a grating spectrometer, but the requirement for input to a fiber spectrometer is more restrictive. While the entrance slit in a grating spectrometer can be opened further to collect more light at the cost of lower resolution, exchanging the single-mode fiber for a few-mode fiber is more complicated. The input light must be spatially incoherent, namely, the light coupled to different spatial modes of the few-mode fiber must have uncorrelated phase, so that the speckle patterns generated by each mode will add in intensity. The transmission matrix then needs to be recalibrated, in order to include the speckle patterns not only from each spectral channel, but also from each spatial mode in the few-mode fiber. This would reduce the operation bandwidth, since the reconstruction algorithm has to find both the spatial profile and the spectral content of the input signal. We therefore expect the fiber spectrometer to be most useful in applications which already require light to be delivered by single-mode fibers. Examples include optical wavemeters, spectral channel monitors used in telecommunications, or fiber based sensors. In addition, spectroscopy techniques which utilize spatial confocality to probe small spatial volumes can use a single-mode fiber for signal collection, enabling integration with the fiber spectrometer. Examples include optical coherence tomography, micro-photoluminescence spectroscopy, and micro-Raman spectroscopy.

**Methods**

We used step-index multimode glass fibers with NA=0.22, a core diameter of 105 um, and a cladding diameter of 125 um. The 100 meter fiber used for the IR spectrometer was a low-OH core Nufern fiber (MM-S105/125-22A) and the 4 cm fiber used for the visible spectrometer was a silica core Thorlabs fiber (FG105LCA). The Nufern fiber introduced attenuation of ~0.1 dB over 100 meters whereas the 4 cm fiber was too short to accumulate any measurable loss. The 100 meter fiber spectrometer was tested using a tunable, near-infrared laser (Agilent 81600B), which was coupled to a single-mode polarization-maintaining fiber, which was spliced to the multimode fiber. The speckle pattern was recorded using an InGaAs camera (Xenics Xeva 1.7-320). The experiments performed using the 100 meter fiber were conducted at least two weeks after the fiber was coiled on the 3" spool to allow time for the initial mechanical relaxation. The 4 cm fiber spectrometer was calibrated using a supercontinuum light source (Fianium WhiteLase SC400-4) coupled through a monochromator (Acton Research Corp. SpectraPro 500) with three gratings (150, 300, or 1200 grooves/mm). A 2 meter long, single-mode polarization-maintaining fiber (Thorlabs P1-488PM-FC-2) was used to couple light to the 4 cm multimode fiber. The speckle pattern was then recorded using a monochrome CCD camera (Allied Vision Manta).

We are grateful to Peter Rakich and Heedeuk Shin for aid in fiber splicing and providing access to their tunable laser. We also thank Kerry Vahala, Doug Stone, and Sebastien Popoff for useful discussions. This work was supported in part by the National Science Foundation under the Grant No. ECCS-1128542.

**References**

1. Xu, Z. *et al.* Multimodal multiplex spectroscopy using photonic crystals. *Opt. Express* **11,** 2126–33 (2003).

2. Kohlgraf-Owens, T. W. & Dogariu, A. Transmission matrices of random media: means for spectral polarimetric measurements. *Opt. Lett.* **35,** 2236–8 (2010).

3. Redding, B., Liew, S. F., Sarma, R. & Cao, H. Compact spectrometer based on a disordered photonic chip. *Nat. Photonics* **7,** 746 (2013).

4. Hang, Q., Ung, B., Syed, I., Guo, N. & Skorobogatiy, M. Photonic bandgap fiber bundle spectrometer. *Appl. Opt.* **49,** 4791–800 (2010).

5. Redding, B. & Cao, H. Using a multimode fiber as a high-resolution, low-loss spectrometer. *Opt. Lett.* **37,** 3384–6 (2012).

6.  Redding, B., Popoff, S. M. & Cao, H. All-fiber spectrometer based on speckle pattern reconstruction. *Opt. Express* **21,** 6584–6600 (2013).

7.  Rawson, E. G., Goodman, J. W. & Norton, R. E. Frequency dependence of modal noise in multimode optical fibers. *J. Opt. Soc. Am.* **70,** 968-976 (1980).

8.  Hlubina, P. Spectral and Dispersion Analysis of Laser Sources and Multimode Fibres Via the Statistics of the Intensity Pattern. *J. Mod. Opt.* **41,** 1001–1014 (1994).

9.  Freude, W., Fritzsche, C., Grau, G. & Shan-Da, L. Speckle interferometry for spectral analysis of laser sources and multimode optical waveguides. *J. Light. Technol.* **LT-4,** 64–72 (1986).

10. Ho, K.-P. & Kahn, J. M. Mode Coupling and its Impact on Spatially Multiplexed Systems in *Opt. Fiber Telecommun. VI* 1–65 (2013).

11. Goodman, J. W. *Speckle Phenomena in Optics*. 141–186 (Roberts & Company, 2007).

12. Okamoto, T. & Yamaguchi, I. Multimode fiber-optic Mach-Zehnder Interferometer and its use in temperature measurement. *Appl. Opt.* **27,** 3085–3087 (1988).

13. Choi, H. S., Taylor, H. F. & Lee, C. E. High-performance fiber-optic temperature sensor using low-coherence interferometry. *Opt. Lett.* **22,** 1814–1816 (1997).

14. Pan, K., Uang, C. M., Cheng, F. & Yu, F. T. S. Multimode fiber sensing by using mean-absolute speckle-intensity variation. *Appl. Opt.* **33,** 2095–2098 (1994).

15. Oraby, O. A., Spencer, J. W. & Jones, G. R. Monitoring changes in the speckle field from an optical fibre exposed to low frequency acoustical vibrations. *J. Mod. Opt.* **56,** 55–66 (2009).

16. Redding, B., Popoff, S. M., Bromberg, Y., Choma, M. A. & Cao, H. Noise analysis of spectrometers based on speckle pattern reconstruction. *Applied Optics* **53**, 410-417 (2014).